\documentclass[aip,jcp,citeautoscript,unsortedaddress]{revtex4-1}

\usepackage{graphicx}
 \usepackage{epstopdf}
\usepackage{dcolumn}%
\usepackage{bm}%
\usepackage{amsmath}
\usepackage[english]{babel}

\begin{document}

\title{Analysis of the excited-state absorption spectral bandshape of oligofluorenes}

\author{Sophia C.\ Hayes}
\email{shayes@ucy.ac.cy}
\affiliation{Department of Chemistry, University of Cyprus, P.O.\ Box 20537, 1678 Nicosia, Cyprus}

\author{Carlos Silva}
\email{carlos.silva@umontreal.ca}
\affiliation{D\'{e}partement de physique et Regroupement qu\'{e}b\'{e}cois sur les mat\'{e}riaux de pointe, Universit\'{e} de Montr\'{e}al,  C.P.\ 6128, Succursale centre-ville, Montr\'{e}al (Qu\'{e}bec), H3C~3J7, Canada}

\date{\today}

\begin{abstract}
We present ultrafast transient absorption spectra of two oligofluorene derivatives in dilute solution. These spectra display a photoinduced absorption band with clear vibronic structure, which we analyze rigorously using a time-dependent formalism of absorption to extract the principal excited-state vibrational normal-mode frequencies that couple to the electronic transition, the configurational displacement of the higher-lying excited state, and the reorganization energies. We can model the excited-state absorption spectrum using two totally symmetric vibrational modes with frequencies 450 (dimer) or 400\,cm$^{-1}$ (trimer), and 1666\,cm$^{-1}$. The reorganization energy of the ground-state absorption is rather insensitive to the oligomer length at 230\,meV. However, that of the excited-state absorption evolves from 58 to 166\,meV between the oligofluorene dimer and trimer. Based on previous theoretical work [Shukla~et~al., Phys.\ Rev.\ B \textbf{67}, 245203 (2003)], we assign the absorption spectra to a transition from the $1B_u$ excited state to a higher-lying $mA_g$ state, and find that the energy of the excited-state transition with respect to the ground-state transition energy is in excellent agreement with the theoretical predictions for both oligomers studied here. These results and analysis permit profound understanding of the nature of excited-state absorption in $\pi$-conjugated polymers, which are the subject of general interest as organic semiconductors in the solid state.
\end{abstract}

\maketitle

\section{INTRODUCTION\label{sec:intro}}

Understanding the electronic structure of $\pi$-conjugated polymers and oligomers is of fundamental importance in chemical physics, and also essential for the design of next-generation polymeric semiconductors for applications in optoelectronics.  
Polyphenylenes possess D$_{2\mathrm{h}}$ symmetry; therefore their electronic-state manifold consists of alternating even- ($A_g$) and odd-parity ($B_u$) states, with the lowest optically-active excited state of $B_u$ symmetry.  Excited-state absorption spectra have been measured in a variety of conjugated polymers by means of ultrafast transient absorption spectroscopy, and have been assigned to transitions to various $A_g$ states coupled to the 1$B_u$ excited state. Subpicosecond transient spectra reported in various substituted PPV derivatives show the presence of a strong transient feature at $\sim 1$\,eV, together with a weaker band lying around 1.5\,eV~\cite{Frolov:2000p4454,Kraabel:2000p4475,Frolov:2002p4443}, assigned to transitions to the $mA_g$ and $kA_g$ states respectively. The occurrence of two distinct photoinduced absorption features has also been observed in the transient spectra of higher optical-gap materials such as polyfluorenes (PF)~\cite{Stevens:2001p55,Xu:2001p4474,Korovyanko:2002p4437,Tong:2007p4171},  polyindenofluorenes (PIF)~\cite{Silva:2000p9} and ladder-type polyparaphenylenes (LPPP)~\cite{Cerullo:1998p4467,Graupner:1998p4470,Zenz:2001p4433}, where the low energy feature lies at $\sim 1.5$\,eV, with a broad shoulder in the range 1.8--2.2\,eV, arising from transitions to the $mA_g$ state and a superposition of exciton and polaron absorptions, respectively.
Numerous theoretical investigations have been undertaken in order to provide a clear understanding of the electronic manifold of conjugated polymers, and, more importantly, to probe the nature of the electronic states, as it relates to experimental findings in polyphenylenes~\cite{Chakrabarti:1999p4458,Dixit:1991p4439,Lavrentiev:1999p4452,Psiachos:2009p3231,Shuai:1992p4446,Shukla:2003p1549,Soos:1991p4434,Wang:2008p1409,Yaron:1996p4466}. In the initial work of Chakrabarti and Mazumdar on even-parity states in polyphenylenes~\cite{Chakrabarti:1999p4458}, the limitations of linear-chain models to describe electronic excitations in such materials were recognized, and consideration of both delocalized and localized benzene molecular orbitals (MOs) was deemed essential in the description of $A_g$ states. This work identified two distinct classes of even parity states; an $mA_g$ state that is more delocalized in character and a new class of $A_g$ states involving both delocalized and localized MOs, which according to the calculations, would account for the two PA features in the experimental work. Later work by Shukla~et~al.\ improved this model through correlated electron calculations in PPP and PPV oligomers that sought to describe the distinction between classes of even-parity states more accurately by including the full basis set of benzene MOs~\cite{Shukla:2003p1549}. These calculations were able to explain qualitatively the occurrence of the two PA bands at the energies observed experimentally, and suggested that the different nature of the high-lying PA band could explain charge separation from this $kA_g$ state. The same group recently developed a correlated-electron theory of the electronic structure and photophysics of interacting chains of conjugated molecules in order to reconcile differences in the photophysical behavior of these materials observed in solutions and thin films~\cite{Wang:2008p1409}. Very importantly, a one-to-one correspondence between the essential states that produce photoinduced absorption features observed in the two phases was found.  

In this manuscript, we explore the nature of even-parity molecular excited states coupled to the 1$B_u$ state through transient absorption spectroscopy, by probing the evolution in visible PA transition energy as a function of oligomer length in oligofluorenes in dilute solution. Investigation of well-defined oligomers leads to a better understanding of the underlying photophysics of polymeric semiconductor materials.  Here, we present an analysis of the vibronic structure in the photoinduced absorption spectrum, where we show that the main early-time PA spectrum in the visible is due to a single electronic transition, and that the molecular reorganization of the excited $A_g$ state increases with increasing oligomer length.  
The experimentally determined transition energies are in excellent agreement with the predictions of Shukla~et~al.~\cite{Shukla:2003p1549}, and based on this comparison, we assign the PA feature to a vibronic progression of the $mA_g \leftarrow 1B_u$ transition. In our approach to investigate the vibronic structure in the absorption and PA spectra, we borrow terminology used in the theory of resonance Raman spectroscopy, specifically the time-dependent formulation of absorption.  This approach provides a rigorous and more intuitive modeling of vibronic structure with valuable information related to the excited-state potential.  Thus, a combination of experimental and computational modeling provides insights on the nature of higher excited states in phenylene-based conjugated polymers as it relates to existing theoretical calculations~\cite{Shukla:2003p1549,Wang:2008p1409}. 

\section{METHODOLOGIES\label{sec:methods}}

\subsection{Experimental Techniques\label{subsec:exp}}

F6 dimer (OF2) and trimer (OF3) (see chemical structures in Fig.~\ref{fig:structures}) were dissolved in \emph{p}-xylene, with concentration of 0.1 and 0.2\,mg/mL respectively, and were investigated by femtosecond transient absorption spectroscopy. The experimental system consists of a home-built multipass amplified all-solid-state Ti:sapphire laser system producing 70\,fs pulses at 1\,kHz centered at 805\,nm, based on the design of Backus et al~\cite{Backus:2001p1545}. The 325\,nm (3.81\,eV) pump beam attenuated to 45\,$\mu$J/cm$^2$ was obtained by quadrupling the 1300-nm output of an OPA (TOPAS, Light Conversion) which was pumped by 600\,$\mu$J of the laser output.  A small portion ($\sim 2$\,$\mu$J/pulse) of the laser output was used to generate single-filament white light continuum in a sapphire window, and was used as the probe light, which was further split into a signal and a reference beam.  The measurement of the change in probe transmission through the sample due to the pump pulse ($\Delta$\emph{T}/\emph{T}) was made by modulating mechanically the frequency of the pump pulse at half the frequency of the laser, and then  delivering both signal and reference beams to a 300-mm spectrometer and detecting a narrow portion of the white-light spectrum with a pair of avalanche photodiodes.  The output of the photodiodes was integrated and sampled by a PXI data acquisition board (National Instruments) on a shot-by-shot basis, and the running average was determined in real time.  Laser shots where the probe pulse intensity deviated from the mean by more than a predetermined value were excluded from the average.  A typical sweep consisted of an average of 2000 shots per point and a typical transient is on the order of 3 sweeps. The transient spectra were compensated for chirp in the white-light continuum by adjusting the delay at the particular probe photon energy automatically through the acquisition program.  Pump-power dependence measurements were performed to ensure linearity of signal at the experimental pump power.

\subsection{Time-dependent Formulation of Absorption\label{subsec:model}}

In order to model the temperature-dependent absorption spectrum, we apply a time-dependent formalism that takes into account homogeneous and inhomogeneous broadening effects, as well as the equilibrium distribution of initial vibrational states at a given temperature $T$. We consider that the system is initially in a vibrational eigenstate, with quantum number $v$, of the ground-state electronic potential energy surface. At time $t = 0$, incident radiation interacts with the transition dipole moment of the molecule to induce a vertical electronic transition. The initial vibrational state finds itself evolving under the influence of the excited-state Hamiltonian $H$ of which it is not an eigenstate, $| v(t) \rangle = \exp(-i H t / \hbar)\,| v \rangle$. In this picture, the absorption spectrum is the Fourier transform of the time-dependent overlap between the initial vibrational state in the ground state and that propagating under the influence of the excited-state Hamiltonian, $\langle v | v(t) \rangle$, dissipated by pure dephasing and population decay~\cite{Myers:1988p3601,Hayes:2001p4472}. The absorption cross section at incident photon energy $E_{ex}$ and temperature $T$ is thus given by the following expression~\cite{Hayes:2001p4472,Philpott:1998p4460}:
\begin{equation}
\sigma(E_{ex},T) = \frac{2 \pi e^2 M^2 E_{ex}}{3 \hbar^2 c n} \sum_{v} P_{v} \int_{-\infty}^{\infty} dE_0\,\theta(E_0)   \int_{-\infty}^{\infty} dt\,\langle v | v(t) \rangle D(t) \exp  \left ( \imath \frac{ (E_{ex} + \epsilon_{v} )}{\hbar} t \right ).
\label{eq:AbsCrossSec}
\end{equation}
In this equation, $M$ is the transition moment, $e$ is the electronic charge, $n$ is the  index of refraction, $c$ is the speed of light, and  $\theta(E_0)$ is the inhomogeneous broadening corresponding to a Gaussian distribution of transition energies $E_0$ between the ground vibrational states in the ground and excited electronic states. $D(t)$ is a dissipative function which accounts for the homogeneous linewidth $\Gamma$, composed of both pure dephasing and population decay, and discussed in more detail later in this section. It often takes the form of a pure exponential function.  We sum over the occupation probabilities  $P_{v}$ of vibrational states with quantum number $v$, determined from Boltzmann statistics: 
\begin{equation}
P_{v} =  \exp \left ( \frac{-v \hbar \omega}{k T} \right ) \left ( 1 - \exp \left ( \frac{- \hbar \omega}{ k T} \right ) \right ),
\label{eq:boltzmann}
\end{equation}
where $\omega$ is the vibrational frequency of the relevant normal mode. The sum over $v$ in equation~\ref{eq:AbsCrossSec} is required due to the population of levels up to $v = 5$ along low-frequency coordinates (345\,cm$^{-1}$ in the case studied here) at 298\,K. 

The details of the vibronic structure of the absorption spectrum are contained in the term $\langle v | v(t) \rangle$. The absorption correlator is a multidimensional function representing the product of overlaps along each vibrational normal coordinate. Assuming coordinate separability, the absorption time correlator becomes the product of overlaps of all normal modes $\jmath$:
\begin{equation}
\langle v | v(t) \rangle = \prod_{\jmath} \langle v_{\jmath} | v_{\jmath}(t) \rangle.
\label{eq:correlator}
\end{equation}
In this model, we adopt the harmonic approximation, in which ground and excited states are modeled as harmonic surfaces with frequency  $\omega_g$ and $\omega_{e}$, respectively~\cite{Yi:1986p4476}. The single-mode overlap function is
\begin{align}
\langle v_{\jmath} | v_{\jmath}(t) \rangle = &\left [ \psi(t) \right ]^{-1/2} \exp \left[\Delta^2 f(t) \right ] \nonumber \\
&\times \left ( (v_{\jmath}!)^2 2^{2v_{\jmath}} \right )^{-1/2} \left [ \alpha(t) \right ]^{2v_{\jmath}}  \nonumber \\
&\times \sum_{k=0}^{v_{}\jmath} \frac{(2k)!}{k!} \eta_{v_{\jmath} k} \left [ \gamma(t) \right]^{k}  \nonumber \\
&\times H_{2(v_{\jmath}-k)} \left[ \lambda f(t) \Delta / \alpha(t) \right].
\end{align}
Here $H_p$ are Hermite polynomials, 
$\Delta$ is the dimensionless displacement  of the potential wells from the equilibrium position,
 and
\begin{eqnarray}
\psi(t) &=& \frac{\omega_{+}^2}{4 \omega_e \omega_g} \left[ 1 - \left( \frac{\omega_{-}}{\omega_{+}} \right )^2 \exp \left(-2i\omega_e t \right) \right], \\
f(t) &=& - \frac{\omega_g \left[1 - \exp \left(-i \omega_e t \right) \right]}{\omega_{+} - \omega_{-} \exp \left(-i \omega_e t \right)}, \\
\alpha(t) &=& \sqrt{\frac{\omega_{-} - \omega_{+} \exp \left(-i\omega_e t \right)}{2\omega_{+} - 2\omega_{-} \exp \left(-i\omega_e t \right)}} , \\
\gamma(t) &=& \frac{2 \lambda^2 + i \left( \lambda^4 - 1 \right) \sin \omega_e t}{2 \lambda^2 - i \left( \lambda^4 - 1 \right) \sin \omega_e t}, \\
\omega_{\pm} &=& \omega_e \pm \omega_g, \\
\lambda &=& \sqrt{\frac{\omega_e}{\omega_g} } \text{, and}\\
\eta_{v_j k} &=& \sum_{q=0}^{2k} (-1)^q 
\left(
\begin{array}{c}
 v_{\jmath}  \\
  2k-q   
\end{array}
\right)
\left(
\begin{array}{c}
 v_{\jmath} \\
  q
\end{array}
\right).
\end{eqnarray}
Furthermore, we define
\begin{equation}
\left(
\begin{array}{c}
  m   \\
  \ell
\end{array}
\right) = \left\{
\begin{array}{l l}
  m! / \left[\ell!(m-\ell)! \right] & \quad \text{when $m \geq \ell$}    \\
  0  & \quad \text{when $m < \ell$} \\
\end{array} \right. .
\end{equation}
These expressions are appropriate for the description of electronic spectra of harmonic molecules in the absence of Dushinsky rotation, and are derived analytically in ref.~\citenum{Yi:1986p4476} (Appendix F).

The most important parameters in the model of the absorption spectrum expressed by equation~\ref{eq:AbsCrossSec} are the ground- ($\omega_{g}$) and excited-state ($\omega_{e}$) vibrational frequencies, the displacement of the potential wells from the equilibrium position ($\Delta$) reflecting structural changes, and the homogeneous linewidth ($\Gamma$).  We will explore here the effect of these parameters on the shape of the absorption spectrum, as these are important for a comprehensive understanding of the spectral evolution observed in the OF2 versus OF3 photoinduced absorption spectra described in Section~\ref{sec:results}. Fig.~\ref{fig:tutorial} presents the calculated absorption spectrum for a \emph{gedanken} molecule with one vibrational mode under various conditions. The effect of varying $\omega_{g}$ and $\omega_{e}$, while keeping all other model parameters constant, is depicted in panel (a).  As the vibronic structure of the absorption spectrum reflects only the excited-state potential energy surface, increasing $\omega_{e}$ with respect to $\omega_{g}$ causes all the vibronic replicas to be blue-shifted. This shift also increases slightly the spectral bandwidth, as well as the resolution of the vibronic structure.  

Differences in the equilibrium electronic structure for the ground and excited states are reflected by configurational displacement of the excited-state potential energy well, $\Delta$.  Displacement of the potential well causes the vertical excitation to access higher states in the vibronic manifold, enhancing the intensity of Frank-Condon factors that correspond to transitions other then 0--0, thus increasing significantly the bandwidth of the spectrum and shifting the maximum to a higher frequency. Fig.~\ref{fig:tutorial}(b) demonstrates the effect of moderate versus small displacement in the potential wells of the above \emph{gedanken} molecule, where decrease in 0--0 intensity and change in the relative intensities of the vibronic replicas are significant.  

The homogeneous linewidth acquires an intuitive meaning in the time-dependent formalism, especially when the dissipation function is exponential in time, i.e.\ $D(t) = \exp({ - \widetilde{\Gamma} t / \hbar })$, where $\widetilde{\Gamma}$ is the dissipation rate constant, which in solution, includes both pure dephasing and population decay. In Fourier space, $D(t)$ in this form corresponds to a Lorentzian spectral bandshape of width $\Gamma$. As the decay due to either molecular collisions or population lifetime becomes faster, the absorption spectrum becomes broader. In Fig.~\ref{fig:tutorial}(c) we observe that as $\Gamma$ is increased five-fold the intensity of the whole spectrum decreases as well as the resolution of the vibronic peaks and the relative intensity between the various transitions.  A similar damping effect is accomplished by including in the calculation an appreciably displaced second mode of low-frequency, as its overlap $\langle v | v(t) \rangle$ gets truncated earlier causing short-time behavior and broadening effects~\cite{Karabunarliev:2000p4468,Myers:1988p3601}.

\section{RESULTS AND ANALYSIS\label{sec:results}}

Fig.~\ref{fig:spectra} presents the ground-state absorption spectra for OF2 and OF3 (panel (a)) along with their respective early-time (1-ps) photoinduced absorption (PA) spectra in the visible region after excitation at 3.81\,eV (panel b). The absorption cross section for the photoinduced absorption process has been determined from the transient absorption data using the Beer-Lambert law in the following form
\begin{equation}
\sigma(E_{ex}) = \frac{-1}{n_{ex} z} \ln \left( 1+ \frac{\Delta T}{T} \left( E_{ex} \right) \right ),
\label{eq:BeerLambert}
\end{equation}
where $n_{ex}$ is the population density of the $1B_u$ state and $z$ is the pathlength through the sample.  The population density (cm$^{-3}$) has been calculated using the ground-state concentration (molecules/cm$^{3}$), the photon flux (photons/cm$^{2}$) and the ground-state absorption cross section (cm$^2$, which represents the probability for a transition). We assign the PA band to the $mA_g \leftarrow 1B_u$ transition, as will be discussed in detail in  Section~\ref{subsec:length_dep}. In both cases a red shift is observed with increasing number of phenylene rings, which is usually attributed to increased $\pi$-conjugation. In contrast to the broad and featureless ground-state absorption, the oligomer PA spectra show vibronic structure that appears to be dependent on chain length.  The chirp-corrected transient spectra presented here capture the asymmetry of the spectra and indicate the difference in the vibronic structure for these two oligomers.  These studies have also indicated (Fig.~\ref{fig:transients}) that temporal evolution of $\Delta$\emph{T}/\emph{T} is very similar for all wavelengths across the PA band and demonstrates that the molecule has relaxed vibrationally within $1B_u$ by 1\,ps.  In view of our goal to probe and understand experimentally the nature of higher excited states in phenylene-based materials within the existing theoretical framework, we sought to investigate the chain-length dependence of the vibronic structure using a rigorous model of the absorption process based on the time-dependent formalism introduced in Section~\ref{subsec:model}.   

In order to model the absorption spectrum, we select key vibrational modes of the molecule that are considerably displaced in the electronic excited state, i.e. totally symmetric modes of the molecule that are coupled to the electronic transition. Usually such information is provided by resonance Raman experiments as only totally-symmetric modes are enhanced on resonance.  However, oligofluorenes are very luminescent materials introducing a dominant background in resonance Raman spectra overwhelming the scattered signal. No Resonance Raman studies have been performed so far on oligofluorenes or polyfluorene, however, off-resonance Raman experiments have been performed in a series of oligofluorenes demonstrating a rich vibrational signature, with little variation in the frequencies as a function of chain length~\cite{Tsoi:2008p4461}.  The most intense mode in those vibrational spectra is the symmetric ring stretching mode at $\sim 1607$\,cm$^{-1}$.  This totally symmetric mode has been shown theoretically to couple strongly to the ground-state electronic transition ($1B_u \leftarrow 1A_g$) in the case of OPVÕs (oligophenylvinylenes)~\cite{Karabunarliev:2000p4468}.  In that study, about 20 modes were included in the calculation of the ground-state absorption spectrum, from which four characteristic modes were the symmetric ring-stretch mentioned above, a high-frequency mode corresponding to the C=C stretch in the vinylic moiety, a longitudinal molecular stretch low-frequency mode, and a librational mode. Ozaki~et~al.\ performed resonance Raman experiments on films of derivatives of PPV and PPE~\cite{Ozaki1997}, where they obtained the vibrational spectrum in the region 1100 to 2200\,cm$^{-1}$.  In this region, the prominent peaks correspond to the two high-frequency modes as used above (ring-C---C stretch and the vinylic C=C stretch). Various C---C stretch modes between monomer units were also observed in this spectral region. 

Fig.~\ref{fig:model_dimer} presents the simulation of the ground-state absorption and the photoinduced absorption spectrum of OF2. In these simulations, the simple harmonic approximation described above was used in which ground and excited states were modeled as harmonic surfaces of either equal or different frequency. 
We explored the possibility for anharmonicity by carrying out the propagation $\langle v | v(t) \rangle$  using the approximate method of Feit and Fleck~\cite{Feit:1982p4426,Feit:1983p4430}, but these simulations did not result in better agreement with the experimental data; therefore the harmonic approximation introduced in Section~\ref{subsec:model} was used in all the modeling presented here. The time-dependent overlaps and corresponding absorption cross section were calculated for every ground-state configuration for which the occupation probability $P_v$ (equation~\ref{eq:boltzmann}) was $\geq 1\times10^{-5}$.  This probability cutoff allowed for inclusion of $>99$\% of the ground-state population. We employed a 0.5-fs time step, and we calculated overlaps up to 500\,fs. In the present study, two modes were used to fit all the spectra.  These were the phenylene symmetric ring stretch and a low frequency mode.  From the spectra of Tsoi and Lidzey~\cite{Tsoi:2008p4461}, the 345\,cm$^{-1}$ mode intensity appeared to be most sensitive to chain length, and thus was the low-frequency mode of choice in our modeling. We infer that this mode corresponds most probably to a ring torsion mode. In addition, previous studies on PPV indicate that  torsional modes are the lowest frequency modes that are effectively coupled to the $\pi$-electron system, confirming the choice of such mode in the simulation~\cite{Pichler:1993p4479}. Previous work by Wu~et~al.\ applied a time-dependent Franck-Condon analysis to simulate the temperature-dependent vibronic structure in the photoluminescence spectra of distyrylbenzene~\cite{Wu2005}.  In this analysis they also used a high and lower frequency mode to simulate the PL spectrum at lower temperatures. As observed in Fig.~\ref{fig:model_dimer}(a), inclusion of this mode improved the fit of the ground-state absorption spectrum, even though the spectrum is quite broad, and helped to minimize the contribution of homogeneous linewidth to the broadening of the spectrum. Inclusion of other vibrational modes of either higher or similar frequency did not improve the fit significantly, therefore all the simulations presented below include only two appreciably displaced vibrational modes. The parameters used in all the simulations are presented in Table~\ref{GroundFitParams}. It is noteworthy that when modeling the ground-state absorption cross-section it was necessary to increase the excited-state frequencies of both modes to 1666 from 1607\,cm$^{-1}$ and to 450 from 345\,cm$^{-1}$ respectively.  This is consistent with the widely accepted view that aromatic conjugated polymers assume a quinoidal form in the electronic excited state, where there is an alternation of single and double carbon-carbon bonds and the ring C---C symmetric stretching mode lies at a higher frequency. 
Even with the addition of a displaced low-frequency mode, the contribution of homogeneous and inhomogeneous broadening is still significant. However, it is difficult to set an upper limit for the broadening or to distinguish between the two effects with the absorption spectrum alone. This can be usually accomplished by simultaneously fitting the absorption spectrum and resonance Raman excitation profiles, as the absolute resonance Raman cross sections depend very sensitively upon the homogeneous linewidth.  However, as already mentioned, oligofluorenes are very luminescent materials, with the photoluminescence background in resonance Raman spectra overwhelming the scattered signal, inhibiting the collection of useful spectra.  

Using the excited-state ($1B_u$) frequencies obtained from the fit to the ground-state absorption spectrum as a starting point for the photoinduced absorption transition, we were able to reproduce well the vibronic structure for OF2 as demonstrated in Fig.~\ref{fig:model_dimer}(b).  In this case, $\omega_g$ and $\omega_e$ remained the same, while the displacement in both modes decreased significantly (more than a factor of two for the low frequency mode).  The latter indicates that transitions to higher excited states do not  distort significantly the electronic configuration.  Calculation of the absorption cross section for the PA process allows a more accurate determination of the transition dipole moment in the simulation.  This is constrained to less than half the value for the ground-state absorption process, indicating a smaller probability for this transition than the $1B_u \leftarrow 1A_g$ absorption. Finally, even though a Gaussian exponential form for $D(t)$ was found to best reproduce the red edge of the ground-state absorption spectrum, a Lorentzian function better reproduced the photoinduced absorption spectrum.

Fig.~\ref{fig:model_trimer} presents the simulation of the absorption and photoinduced absorption spectrum of OF3.  The absorption spectrum for the trimer carries a larger bandwidth than the dimer rendering it insensitive to the number of modes included in the calculation.  However, two modes were used in the fit for consistency with the dimer simulations.  Similar to the case of OF2, good agreement with the ground-state absorption spectrum required the increase in the frequencies of both high and low vibrational modes.  The displacements in both modes were essentially identical to the values used for the dimer, again indicating the transition to a quinoidal form of the electronic configuration.  In contrast, the PA spectrum for OF3 is quite different from OF2.  The bandwidth of the spectrum is wider, and the 0--1 transition is more enhanced in the trimer.  This evolution in vibronic structure with chain length is reproduced well with a significant increase in the displacement of both modes compared to the dimer, with further stiffening of both vibrational modes (Table~\ref{GroundFitParams}).  Specifically, an increase in the symmetric ring stretch frequency to 1740\,cm$^{-1}$, and an increase in the torsional mode frequency to 420\,cm$^{-1}$ was necessary to reproduce the vibronic spacing in the spectrum.   In the PA spectrum of OF3, another feature is visible after the 0--1 transition that appears to be part of the vibronic progression.  However, both harmonic and anharmonic potentials for the symmetric ring stretch mode, which seems to be the dominant mode in these spectra, were not able to reproduce this peak.   Similar to OF2, the transition dipole moment is reduced to about half that for the ground state absorption.  It is interesting to note that for both ground- and excited-state absorption the transition dipole moment is statistically larger in OF3 compared to OF2, in agreement with previous studies on oligofluorenes~\cite{Schumacher:2009p4165}.  

\section{DISCUSSION\label{sec:disc}}

\subsection{Vibronic Analysis of Photoinduced Absorption Spectra\label{subsec:vibronic_analysis}}

The vibronic structure that is evident in the photoinduced absorption (PA) spectra of the oligofluorenes has not been discussed extensively in the literature.  Vibronic structure associated with the excited-state transition has been observed in the PA spectrum of mLPPP and a five-ring oligomer of analogous structure (5LOPP)~\cite{Cerullo:1998p4467,Gadermaier:2002p4428}, but it has not been analyzed. Simulation of the vibronic structure in PA spectra provides an insight on the nature of the optically-accessible higher even-parity excited states in oligofluorenes through the parameters that describe their electronic potential. In the present study, we find that a single electronic transition with evolution along two normal-mode coordinates can reproduce the observed spectrum adequately in both oligomers.  However, the details of this transition differ in the two materials.  

The most striking difference observed in the experimental spectra is the larger enhancement of the 0--1 vibronic peak in the trimer. Following the discussion of model parameters in Section~\ref{subsec:model} and the simulations presented in Fig.~\ref{fig:tutorial}, we interpret this in terms of increased displacement of the equilibrium position in the $mA_g$ state relative to the $1B_u$ state, which reflects the greater distortion in the molecular conformation of the trimer.  Comparison of the displacements and the frequencies of the vibrational modes obtained upon transition to the $1B_u$ state to the ones obtained for transition to the $mA_g$ state is instructive.  Both oligomers display essentially the same change in structure upon excitation to the $1B_u$ state, with stiffening of both vibrational modes.  Increase in the frequency of the symmetric ring C---C stretch to 1666\,cm$^{-1}$ can be related to increase in the double-bond character of this bond that accompanies quinoid-like structures. It is known that the excited-state configuration of
various conjugated molecules such as PPP or PPV is quinoidal~\cite{Bredas:1982p4447,Karabunarliev:2000p4468,Pichler:1993p4479}. This is supported by recent time-dependent density functional theory calculations of terfluorene (no side chains) by Beenken et al., which clearly show that the transition from the ground electronic state to the lowest exciton changes the electronic configuration from benzoid-like to quinoidal~\cite{Beenken:2008p4462}.  The particular frequency of the mode is close to the frequency observed experimentally for the C---C stretch in quinoidal systems~\cite{Kobryanskii:1998p4429}.  The actual frequency depends on the details of each chemical system.  In addition, stiffening of the ring torsion/rotation mode can also be explained from the hindrance to rotation introduced by the double-bond formation between fluorene monomers. Theoretical calculations performed by Karabunarliev~et~al.\ of absorption and fluorescence spectra of various conjugated molecules including oligophenylenevinylenes of various lengths, using a rigorous Franck-Condon analysis, demonstrated stiffening of ring torsional modes in the $1B_u$ state~\cite{Karabunarliev:2000p4468}; however, these calculations do not predict an effective change in the excited state ring C---C frequency while this is observed for the C=C stretch.  In a system without a vinylic bond, we expect that stiffening will be observed in the ring C---C mode. Further excitation to the $mA_g$ state does not change the quinoidal structure of the dimer significantly, as the mode frequencies remain the same and the displacement is small, in contrast to the case of the trimer.  The increased displacement and frequencies in the trimer, though,  imply that the molecular reorganization of the $mA_g$ state increases upon increasing monomer units from two to three.

The reorganization energy $\lambda$ can be viewed as the energy difference between two geometrically different structures in the electronic excited state; the structure at the equilibrium geometry of the ground state after a vertical transition, and that at the equilibrium geometry of the excited state.  This energy includes contributions from the internal reorganization of the molecule and the reorganization of the solvent environment to adjust to changes in the electronic configuration of the solute.  Reorganization energies can be calculated by the following expression~\cite{Myers:1996p4167}, given the excited- ($\omega_{ej}$) and ground-state ($\omega_{gj}$) frequencies, and the displacement ($\Delta_{j}$) along the dimensionless normal coordinate of the ground state of mode $j$:  
 \begin{equation}
\lambda_j = \left( \frac{\omega_{ej}^2}{\omega_{gj}} \right) \frac{\Delta^2}{2}.
\label{eq:Reorg}
\end{equation}
Such information can be drawn experimentally from resonance Raman intensity analysis, where simultaneous modeling of Raman excitation profiles and the absorption spectrum is performed~\cite{Myers:1988p3601}, which is not feasible in the present study.  Excited-state resonance Raman spectra are even more rare, which, in addition to the fluorescence background problem, makes experimental results sparse.  Therefore, even if the simulation of the absorption and photoinduced absorption spectra here is simplified to include only two totally symmetric modes, calculation of the reorganization energies using equation~\ref{eq:Reorg} can still provide useful and self-consistent information.  The calculated mode-specific reorganization energies for both transitions and for both oligomers are shown in Table~\ref{ReorgEnerg}.  Consistent with the observations in the spectra and the simulations performed, the reorganization energy for the ground-state absorption is similar for both oligomers, where both molecules are expected to undergo a large configurational change from aromatic to quinoidal form. The total reorganization energy is consistent with the $\lambda$ obtained from experimental measurements of the vertical transition and the $E_0$ absorption in oligofluorenes~\cite{Schumacher:2009p4165}. However, the value of $\lambda$ calculated here for the $mA_g \leftarrow 1B_u$ transition is a factor of three larger in the trimer for both vibrational modes.  Reorganization energies have not been, to the best of our knowledge, reported in the literature for higher even-parity states of conjugated oligomers or polymers. Similarly, calculations of charge densities of such states that could aid in the assignment of the larger reorganization energy in the trimer have not been performed.  Cornil~et~al.\ reported experimental and theoretical studies on PPV oligomers, where they observed a nearly constant Huang-Rhys factor for absorption (the only difference was between $n=2$ and $n=3$ oligomers)~\cite{Cornil1995425, Cornil1997139}.  On the other hand, the PL spectra demonstrated decreasing $\Delta$ as a function of chain length, as observed from increased intensity in the 0--0 vibronic band. In PPV oligomers, the ground state is expected to be more planar than in the case of fluorene oligomers, where a single bond links monomer units.  Also, the vibronic spacing in PL is smaller than in absorption, which is probably due to coupling of different modes to the vibronic transition~\cite{Karabunarliev:2000p4468}.   Chain-length dependence studies of the absorption and PL spectrum in oligofluorenes were performed in frozen solutions at 77\,K, where a similar trend as in PPV is observed~\cite{Wasserberg:2005p4481, Schumacher:2009p4165}. In the present study, essentially no difference in the reorganization energy is observed for the ground state at room temperature for the two oligomers, where the vibronic structure is obscured due to conformational disorder of the inter-ring dihedral angles between repeat units (giving rise to inhomogeneous broadening~\cite{Wasserberg:2005p4481}). In the case of absorption to higher excited states, not much is known about the geometric structure of oligofluorenes or PF.
 We believe that increasing in the chain length increases torsional disorder in solution, thus excitation to a state where further planarization of the molecule occurs should involve, on average, a greater reorganization energy, both in terms of internal and solvent  reorganization. This is supported by the study in ref.~\citenum{Hutchison:2005p4436}, where the authors suggested that geometrical changes such as bond length alternation and inter-ring dihedral angles changes can have a contribution to reorganization energy, along with oligomer length and increased delocalization.  

\subsection{Nature of the excited-state transition\label{subsec:length_dep}}

Ultrafast transient absorption measurements in OF2 and OF3 performed in this study revealed a transition peaked around 1.95 and 1.8\,eV, respectively (Fig.~\ref{fig:spectra}), with origin $E_0$ at 1.87 and 1.66\,eV (Table~\ref{GroundFitParams}). We assign this absorption to the $mA_g \leftarrow 1B_u$ intrachain transition according to previous literature on polyfluorenes in thin films~\cite{Stevens:2001p55,Xu:2001p4474,Korovyanko:2002p4437}. Transient absorption work on polyfluorene (PF) in solution~\cite{Korovyanko:2002p4437} demonstrated a photoinduced absorption band around 1.7\,eV, assigned to an intrachain exciton as measured here.  From the $E_0$ transition energies in Table~\ref{GroundFitParams}, and the data reported in ref.~\citenum{Korovyanko:2002p4437}, we infer that the transition energy varies linearly as a function of inverse number of phenyl rings (Fig.~\ref{fig:lengthdep}).  We observe a similar trend for the ground-state absorption of these materials.  This is in agreement with planarization of the excited state geometry towards a quinoidal structure  with increased excited-state delocalization~\cite{Karabunarliev:2000p4468}, which is in accord with the simulations presented in this study.
 
Correlated electron calculations have been performed by Shukla~et~al., in which different classes of two-photon ($A_g$) states have been computed explicitly using a complete set of benzene molecular orbitals~\cite{Shukla:2003p1549}. Four spectral features were calculated arising from these states, which can be grouped into three categories. The calculated PA spectra were labelled as features I, II and III which were grouped together, and IV.  The first group of states is described as a superposition of $1e$--$1h$ $d_1\rightarrow d_1^{*}$  single excitation (i.e.\ excitation from an innermost delocalized valence band to a delocalized innermost conduction band, bonding to antibonding MO) and $2e$--$2h$ ($d_1\rightarrow d_1^{*}$)$^2$ double excitation, with the former carrying most of the weight.  Feature I is then assigned to the $2A_g$ state which is just above the $1B_u$ state, and calculated to appear around 10\% of the $1B_u$ energy. The second group of states is represented by features II and III in the calculated PA spectrum, and show a strong contribution from $2e$--$2h$ ($d_1 \rightarrow d_1^{*}$)$^2$ double excitation.  A weak but nonzero contribution is also observed from ($d_1\rightarrow l^{*}$)$^2$ and ($l \rightarrow d_1^{*}$)$^2$, where $l$ represents a localized MO. In addition, different kinds of single excitations occur with subtle differences in their relative contributions. However, the overall nature of the excitations that constitute bands II and III is similar. These bands have been assigned to transitions from the $1B_u$ to the $mA_g$ state, and have been calculated to occur around 0.4 to 0.6 times the $1B_u$ energy.  Finally, a third class of states exists, with strong contributions from $(d_1 \rightarrow l^{*})^2$, which appears as Feature IV in the calculated PA spectra, with an energy 0.6 to 0.8 times that of $1B_u$. The eigenstate that corresponds to this transition is found to be qualitatively different than the lower $A_g$ states and has been assigned to the $kA_g$ state.  The scaled energies obtained from the calculations of Shukla~et~al.\ for the salient features of the PA spectra have been compared to experimental results for PPV derivatives as well as for higher bandgap materials such as m-LPPP, with very close agreement, when confinement effects are taken into consideration on going from the polymer to the oligomer length. 
 
 Oligofluorenes and polyfluorene are high bandgap materials. Ultrafast transient absorption spectra of polyfluorene show PA bands at around 0.5, 1.6 and 1.8-2.4\,eV~\cite{Xu:2001p4474,Korovyanko:2002p4437,Tong:2007p4171}. Scaling of these energies with respect to the $1B_u$ energy reproduces closely the three spectral regions discussed above, assigned to $2A_g$, $mA_g$ and $kA_g$ states, along with a contribution from polaron pairs for the latter absorption band in the solid state. The $E_0$ transition energies for the excited-state transitions of the oligofluorenes reported in Table~\ref{GroundFitParams}, normalized over the ground-state $E_0$ values reported in the same table, result in $\sim 0.5$ for both materials. We infer the same ratio for PF from  the PA2 band in the transient photoinduced absorption spectrum in ref.~\citenum{Korovyanko:2002p4437}. These are plotted in Fig.~\ref{fig:lengthdep}. Comparison of these ratios with the calculated PA spectra by Shukla~et~al.~\cite{Shukla:2003p1549} clearly indicates that the experimentally observed features correspond to features II and III, which are assigned to the $mA_g \leftarrow 1B_u$ transition. This transition is characterized by strong contributions from delocalized states, which agrees with the significant red-shift observed for both the ground-state and excited-state absorption spectra with increasing chainlength ($\sim 0.5$ and 0.3\,eV from the dimer to the polymer, respectively). This eliminates the possibility that the PA features observed here are related to excitation to the $kA_g$ state, since the latter involves significant contributions from localized orbitals and is expected to show a much weaker chain-length dependence. Assignment of the spectral signature observed here to transition to the $mA_g$ state is also consistent with the chainlength independence of the $E/E(1B_u)$ ratio for feature II as observed in the theoretical calculations for oligomers of PPP and PPV~\cite{Shukla:2003p1549}.

\section{SUMMARY AND CONCLUSIONS\label{sec:conclusions}}
In this manuscript we presented ultrafast photoinduced absorption (PA) measurements of the excited state absorption in two oligofluorenes (OF2 and OF3) in dilute solution.   The photoinduced absorption spectra are observed at 1.95 and 1.8\,eV, respectively, and are assigned to the $mA_g \leftarrow 1B_u$ transition. These transition energies taken along with the PA spectrum of polyfluorene in solution reported in the literature~\cite{Korovyanko:2002p4437} demonstrate a linear dependence of the absorption maximum vs. inverse chainlength, indicating a transition to a delocalized state. The PA spectra measured in this study demonstrate distinct vibronic structure that differs between the two oligomers. We analyzed this structure using the time-dependent formalism of absorption in order to gain insight on the nature of higher even-parity excited states in conjugated molecules.  In this model, we use two moderately displaced totally symmetric modes coupled to the electronic transition to reproduce the vibronic progressions. This analysis revealed that even though both oligomers appear to assume similar quinoidal structures in the $1B_u$ state, further excitation of the trimer to an $A_g$ state requires greater reorganization energy compared to the dimer, possibly due to greater distortion of the longer chain.  In addition, comparison of the ratio $E_0$(PA)/$E_0$($1B_u$) of the transition energies obtained from the analysis of the spectra to the theoretically calculated PA features in the work of Shukla et al.~\cite{Shukla:2003p1549} confirm the assignment of the transition to $mA_g \leftarrow 1B_u$, which is described theoretically to have contributions from delocalized states.

\begin{acknowledgments}

\end{acknowledgments}
CS acknowledges funding from the Natural Sciences and Engineering Research Council (NSERC) and from the Canada Research Chair in Organic Semiconductor Materials. SCH acknowledges funding from the University of Cyprus for collaboration visits to Montreal. We thank Cl\'ement Daniel and Sebastian Westenhoff for their assistance in data acquisition.

%


\newpage
\cleardoublepage

 \begin{table}[H] 
 \caption{Model parameters for the ground-state ($1B_u \leftarrow 1A_g$) and excited-state ($mA_g \leftarrow 1B_u$) absorption spectra of OF2 and OF3 using the time-dependent formalism described in Section~\ref{subsec:model}. Parameter uncertainties are determined by visual comparison of the model results and the experimental data.  \label{GroundFitParams}}
 \begin{ruledtabular}
 \begin{tabular}{c c c c c c c c c}
      & transition & $E_0$\,(cm$^{-1}$) &$\omega_g$\,(cm$^{-1}$) &$ \omega_e$\,(cm$^{-1}$) & $\Delta$ & $M$\,(\AA) & $\Gamma$\,(cm$^{-1}$) & $\theta$\,(cm$^{-1}$) \\
          \hline
 OF2 & $1B_u \leftarrow 1A_g$ & $28200 \pm 50$ & 345\footnotemark[1] & $450 \pm 10$ & $1.4 \pm 0.1$ & $2.45 \pm 0.01$ & $400 \pm 70$ & $500 \pm 100$ \\
 \cline{4-6}
  & & $[3.496 \pm 0.006$\,eV] & 1607\footnotemark[1] & $1666 \pm 20$ & $1.22 \pm 0.03$& & & \\
  \cline{2-9}
   & $mA_g \leftarrow 1B_u$ & $15100 \pm 50$ & $450\footnotemark[2] $ & $450 \pm 20$ & $0.55 \pm 0.1$ & $1.03 \pm 0.01$ & $340 \pm 10$ & $370 \pm 10$ \\
    \cline{4-6}
   & & [$1.872 \pm 0.006$\,eV] & 1666\footnotemark[2] & 1666\footnotemark[3] & $0.69 \pm 0.03$ & & & \\
   \hline
  OF3 & $1B_u \leftarrow 1A_g$& $26500 \pm 50$ & 345\footnotemark[1] & $400 \pm 20$ & $1.4 \pm 0.1$ & $2.54 \pm 0.01$ & $500 \pm 50$ & $500 \pm 50$ \\
  \cline{4-6}
   & & [$3.286 \pm 0.006$\,eV] & 1607\footnotemark[1] & $1666 \pm 15$ & $1.26 \pm 0.02$& & & \\
     \cline{2-9}
 & $mA_g \leftarrow 1B_u$ & $13400 \pm 50$ & 400\footnotemark[2] & $420 \pm 30$ & $1.0 \pm 0.2$ & $1.33 \pm 0.01$ & $230 \pm 10$ & $450 \pm 10$ \\
    \cline{4-6}
  & & [$1.661 \pm 0.006$\,eV] & 1666\footnotemark[2] & $1740 \pm 20$ & $1.10 \pm 0.05$ & & & \\
 \end{tabular}
\end{ruledtabular}
\footnotetext[1]{The $1B_u \leftarrow 1A_g$ ground-state vibrational frequencies were taken from reference~[Lidzey].}
\footnotetext[2]{The $mA_g \leftarrow 1B_u$ ground-state vibrational frequencies were the excited-state frequencies determined for the $1B_u \leftarrow 1A_g$ transition.}
\footnotetext[3]{The agreement of the model with the data is not sensitive to this parameter; we therefore do not report an uncertainty.}
 \end{table}
 
  \begin{table}[H] 
 \caption{Total reorganization energies $\lambda_{\text{tot}}$ determined from equation~\ref{eq:Reorg} and Table~\ref{GroundFitParams}.\label{ReorgEnerg}}
 \begin{ruledtabular}
 \begin{tabular}{c c c c c }
        & transition & $\omega_{gj}$\,(cm$^{-1}$) & $\lambda_j$\,(meV) & $\lambda_{\text{tot}}$\,(meV)  \\
          \hline
 OF2 & $1B_u \leftarrow 1A_g$ & 1607& 159 & 230 \\
  & & 345 & 71& \\
  \cline{2-4}
   & $mA_g \leftarrow 1B_u$ & 1666 & 48 &58 \\
 & & 450 & 10 & \\
   \hline
  OF3 & $1B_u \leftarrow 1A_g$ & 1607 & 170&226 \\
 & & 345 & 56& \\
     \cline{2-4}
 & $mA_g \leftarrow 1B_u$ & 1666 &  139&166 \\
 & & 400 & 27& \\
      \end{tabular}
\end{ruledtabular}
 \end{table}
 
 \newpage
 \cleardoublepage
 \section*{FIGURES}

 \begin{figure}[h!]
%
\caption{Chemical structures of OF2 and OF3.}
\label{fig:structures}
\end{figure}

\begin{figure}[h!]
 \caption{Effects of varying the vibrational ground- and excited-state vibrational frequencies (\textit{a}), excited-state displacement (\textit{b}), and homogeneous linewidth (\textit{c}) in a one-mode model of the absorption spectrum as described in Section~\ref{subsec:model}. In panel (a) the solid line represents the calculation where the frequency is kept at 1607\,cm$^{-1}$ for both ground and excited state, while the dotted line corresponds to different ground (1607\,cm$^{-1}$) and excited state frequencies (1750\,cm$^{-1}$).  In panel (b) the frequencies are kept the same at 1666\,cm$^{-1}$. The solid line represents the spectrum obtained with a small displacement ($\Delta = 0.65$) and the dotted line the case with moderate displacement ($\Delta = 1.2$).  In panel (c) all the parameters are kept the same except for the homogeneous linewidth.  The dotted line represents the effect of a five-fold increase in the $\Gamma$, while the grey line uses a second mode at low frequency (450\,cm$^{-1}$) to reproduce the broadening of the spectrum.  All other parameters in the calculations are the same for all the calculations.\label{fig:tutorial}}
\end{figure}

 \begin{figure}[h!]
\caption{(\textit{a}) Ground-state absorption spectra of OF2 (closed triangles) and OF3 (open circles). (\textit{b}) Corresponding excited-state absorption spectra.\label{fig:spectra}}
\end{figure}

\begin{figure}[h!]
 \caption{Absorption transients of OF2 (\textit{a}) and OF3 (\textit{b}) at various probe photon energies, indicated in the insets.\label{fig:transients}}
\end{figure}
 
 \begin{figure}[h!]
 \caption{Simulation of the ground-state (\textit{a}) and excited-state (\textit{b}) absorption spectra of OF2. The parameters used in the simulation are reported in Table~\ref{GroundFitParams}.\label{fig:model_dimer}}
\end{figure}

 \begin{figure}[h!]
 \caption{Simulation of the ground-state (\textit{a}) and excited-state (\textit{b}) absorption spectra of OF3. The parameters used in the simulation are reported in Table~\ref{GroundFitParams}.\label{fig:model_trimer}}
\end{figure}

 \begin{figure}[h!]
 \caption{Transition energy ($E_0$) as a function of inverse number of phenyl rings (1/$n$) constituting the oligomer (left axis). The data for the dimer and trimer are obtained from Table~\ref{GroundFitParams}. The data point at 1/$n$ = 0 (with the large $x$-axis incertitude) corresponds to a polyfluorene derivative, taken from ref.~\citenum{Korovyanko:2002p4437}. Also shown is the ratio of excited-state to ground-state transition energies (right axis).  \label{fig:lengthdep}}
\end{figure}
 
 \newpage
 \cleardoublepage
 \begin{center}
\includegraphics[scale=1]{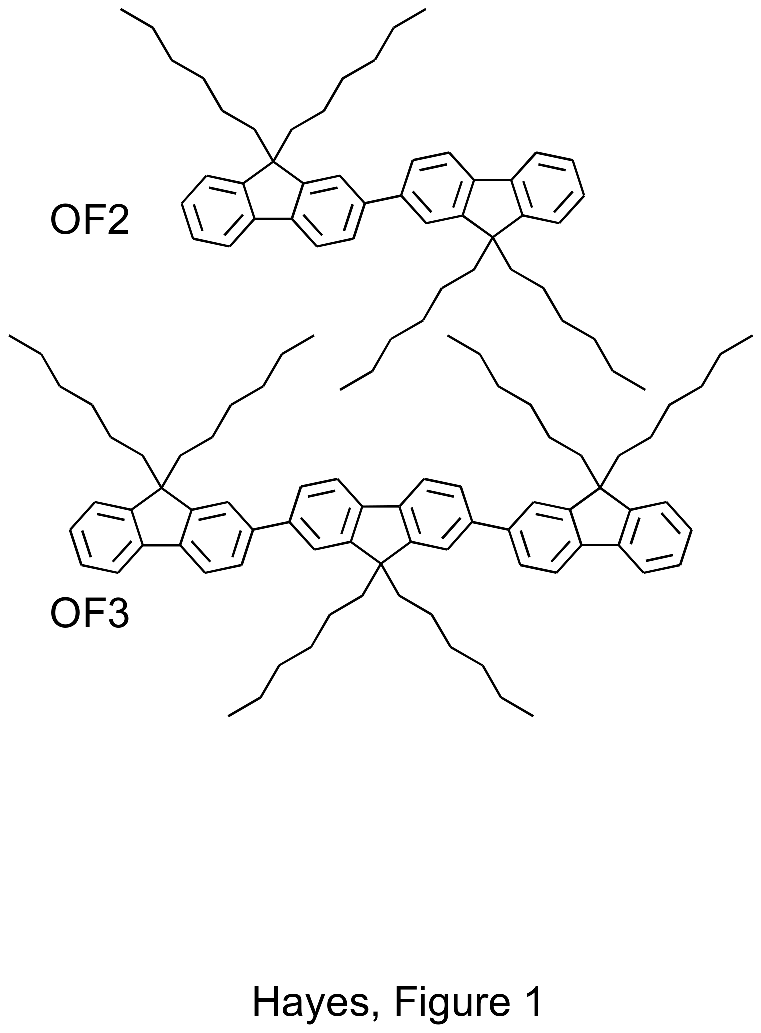} 
\end{center}

 \newpage
  \begin{center}
\includegraphics[scale=1]{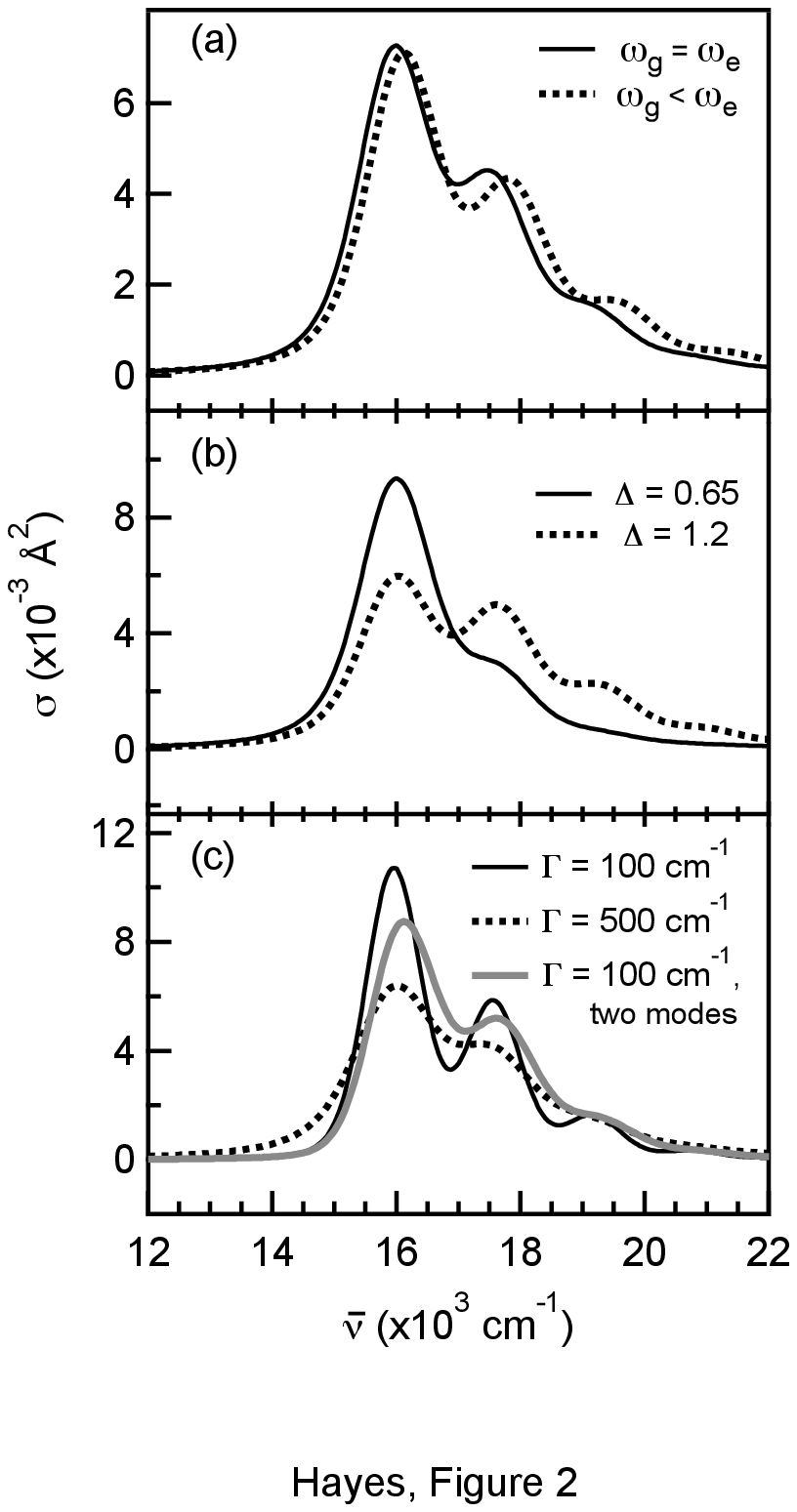} 
\end{center}

\newpage
 \begin{center}
\includegraphics[scale=1]{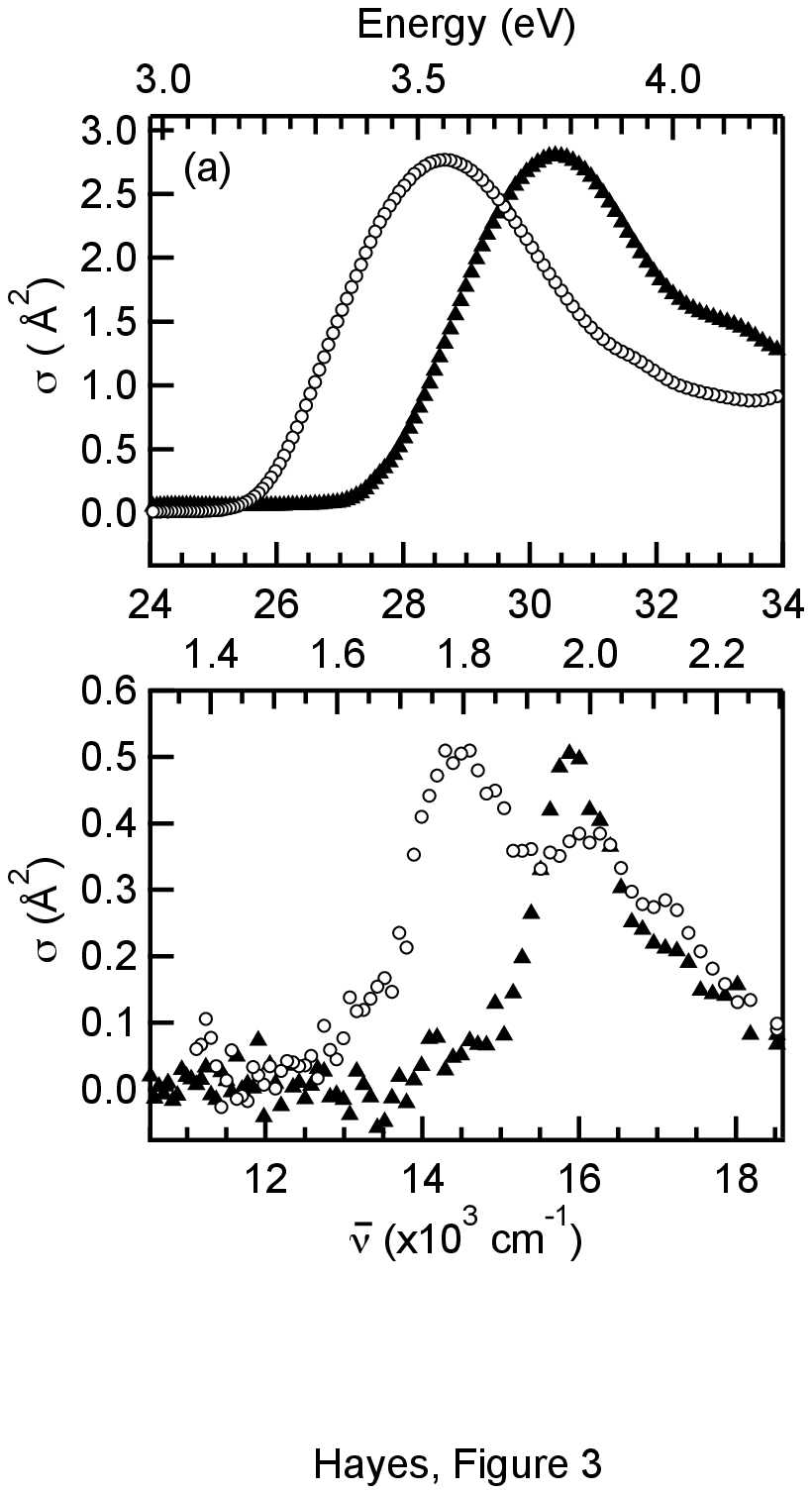} 
\end{center}

\newpage
 \begin{center}
\includegraphics[scale=1]{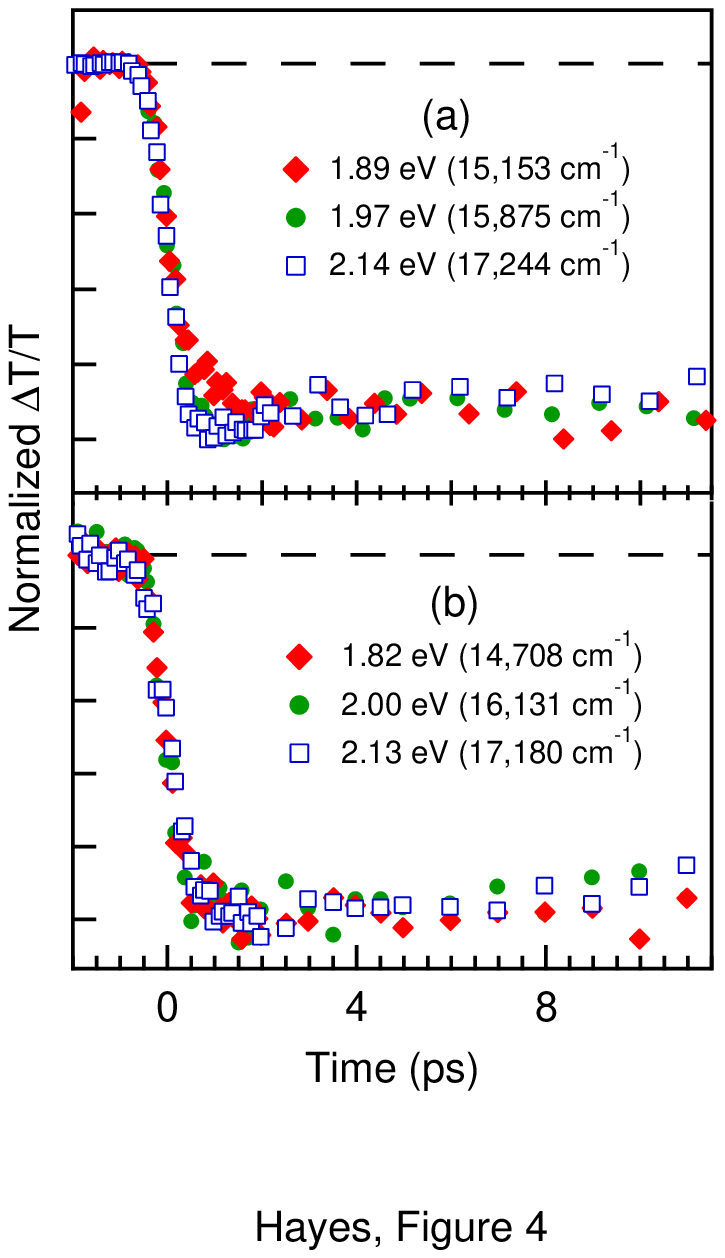} 
\end{center}

\newpage
 \begin{center}
\includegraphics[scale=1]{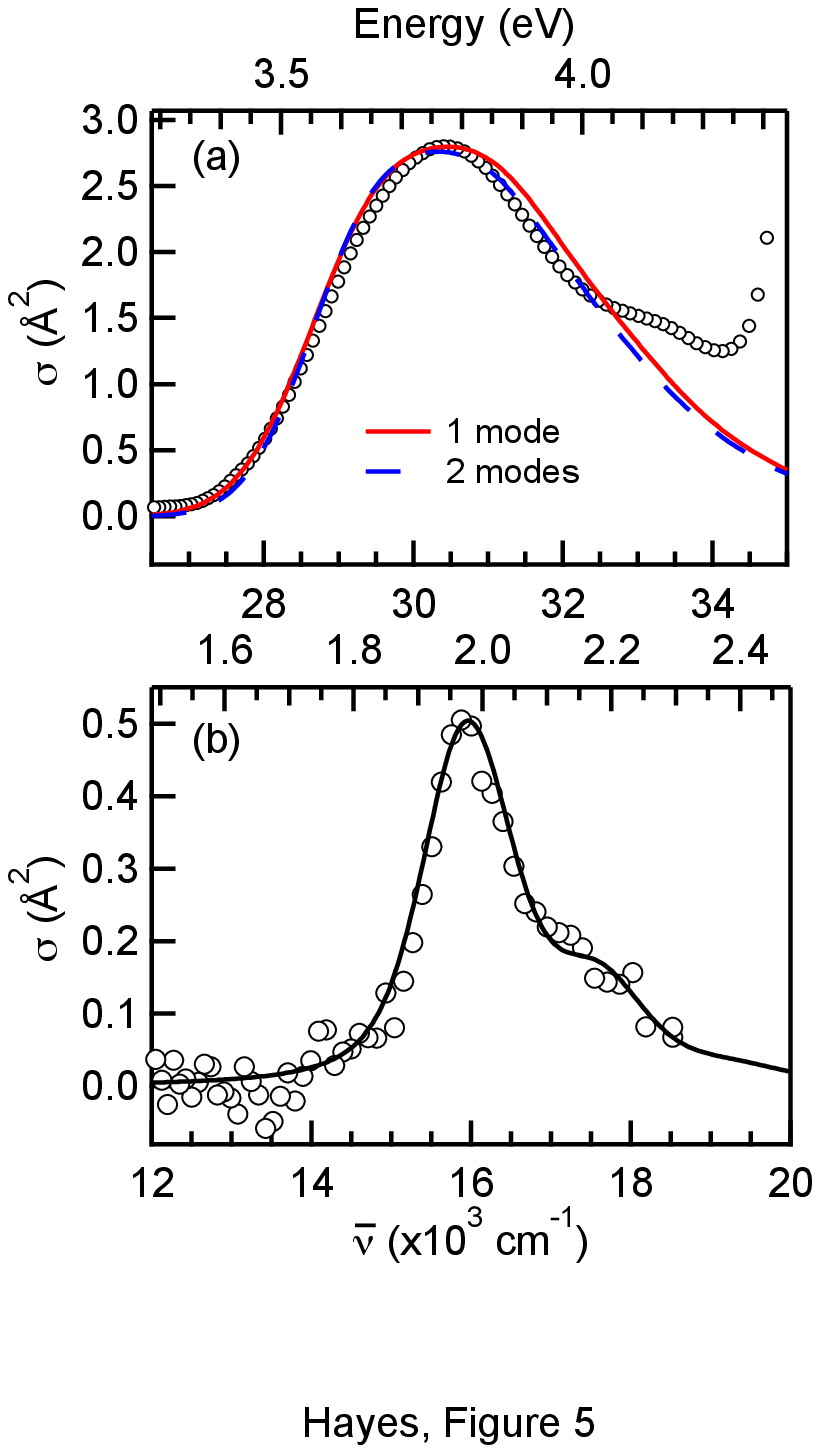} 
\end{center}

\newpage
 \begin{center}
\includegraphics[scale=1]{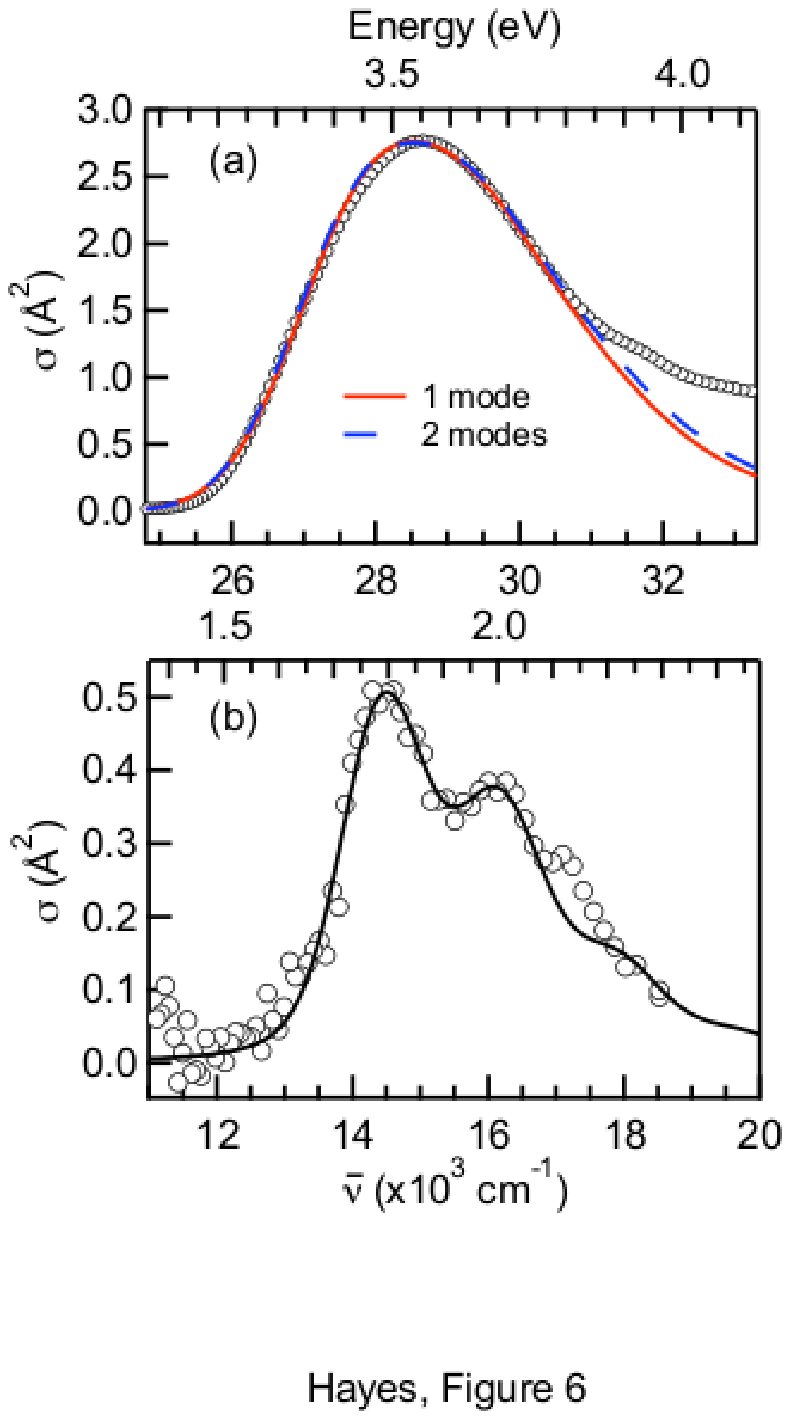} 
\end{center}

\newpage
 \begin{center}
\includegraphics[scale=1]{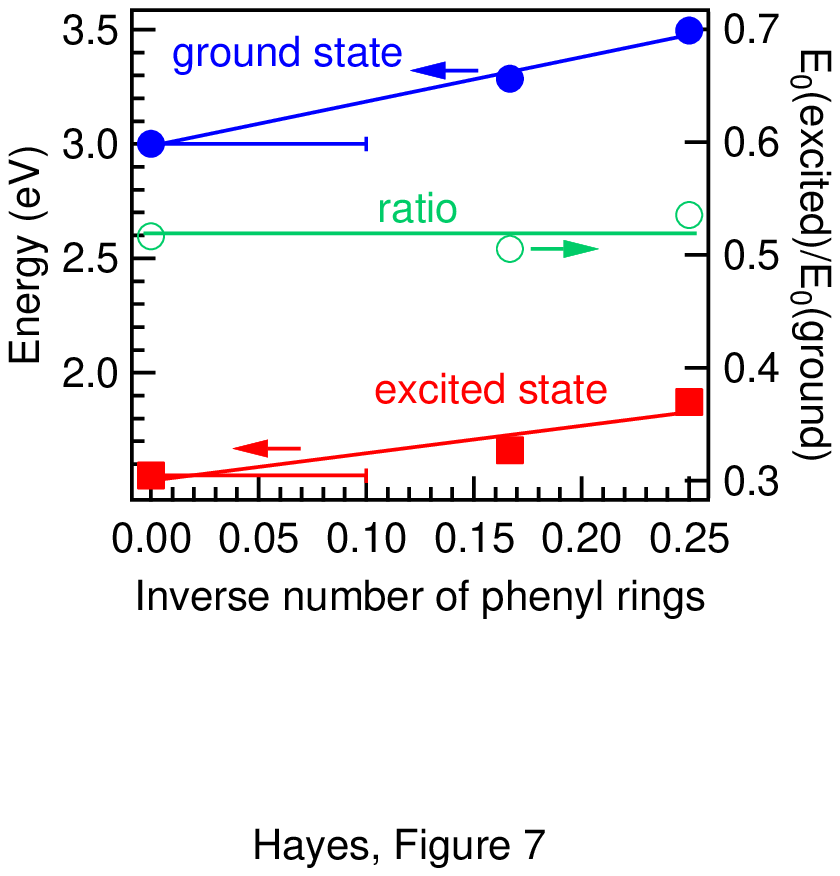} 
\end{center}

\end{document}